\documentclass[referee]{raa}            % referee version: for submission

\usepackage{graphicx,times}             %for PS/EPS graphics inclusion, new
\usepackage{natbib}
\usepackage{pdflscape}
\usepackage{multirow}
\usepackage{array}
\usepackage{threeparttable}
\usepackage{booktabs}
\usepackage{indentfirst}
\usepackage{color}
\usepackage{amssymb,amsmath}
\usepackage{lineno}
\begin{document}
   \title{Is the enhancement of type II radio bursts during CME interactions related to the associated solar energetic particle event?
%\footnotetext{$*$ Supported by the National Natural Science Foundation of China.}
}
%   \subtitle{I. Place Your Subtitle Here}

   \volnopage{Vol.0 (201x) No.0, 000--000}      %%preserved for Editor. DOn't remove!
   \setcounter{page}{1}          %%starting page, preserved for Editor. DOn't remove!

   \author{Liu-Guan, Ding
      \inst{1,2}
   \and Zhi-Wei, Wang
       \inst{2}
   \and Li, Feng
       \inst{3}
   \and Gang, Li
      \inst{4}
   \and Yong, Jiang
       \inst{2}
   }

   \institute{School of Physics and Optoelectronic Engineering, Nanjing University of Information Science and Technology, Nanjing 210044, China; {\it dlg@nuist.edu.cn}\\
      \and    Institute of Space Weather, Nanjing University of Information Science and Technology, Nanjing 210044, China;\\
   \and    Key Laboratory of Dark Matter and Space Astronomy, Purple Mountain Observatory, Chinese Academy
of Sciences, 210008 Nanjing, China \\
   \and    Department of Space Science and CSPAR, University of Alabama in Huntsville, AL, 35899, USA \\
     }

   \date{Received~~2018 month day; accepted~~2018~~month day}

%\linenumbers*[1]

\abstract{
We investigated 64 pairs of interacting-CME events identified by the simultaneous observations of SOHO and STEREO spacecraft from 2010 January to 2014 August, to examine the relationship between the large SEP events in the energy of $\sim25-\sim60$MeV and the properties of the interacting CMEs.
We found that during CME interactions the large SEP events in this study were all generated by CMEs with the presence of enhanced type II radio bursts, which also have wider longitudinal distributions comparing to events with the absence of type II radio burst or its enhancement (almost associated with small SEP events).
It seems that the signature of type II radio bursts enhancement is a good discriminator between large SEP and small or none SEP event producers during CME interactions.
The type II radio burst enhancement is more likely to be generated by {CME interactions}, with the main CME having larger speed ($v$), angular width (WD), mass ($m$) and kinetic energy ($E_k$), that taking over the preceding CMEs which also have higher $v$, WD, $m$ and $E_k$, than those preceding CMEs in CME pairs missing the type II radio bursts or enhancements.
Generally, the type-II-enhanced events typically have higher values of these properties than that of non-type-II or non-type-II-enhanced cases for both the main and the preceding CMEs.
Our analysis also revealed that the intensities of associated SEP events
%, especially for large SEP events,
correlate negatively with the intersection height of the two CMEs. Moreover, the overlap width of two CMEs is typically larger in type-II-enhanced events than in non-type-II or non-type-II-enhanced events.
Most of type-II-enhanced events and SEP events are coincidentally and almost always made by the fast and wide main CMEs that sweeping fully over a relatively slower and narrower preceding CMEs.
We suggest that a fast CME with enough energy completely overtaking a relatively narrower preceding CME, especially in low height, can drive more energetic shock signified by the enhanced type II radio bursts. The shock may accelerate ambient particles (likely provided by the preceding CME) and lead to large SEP event more easily.
}

   \authorrunning{L.G. Ding, et al}            %author_head in even pages
   \titlerunning{Type II enhancement during CME interaction and SEPs}         % title_head in odd pages

   \maketitle
%% The author head (on even pages) and the title head (on odd pages) will be
%% automatically extracted from \author{} and \title{}. Whenever the title is too long,
%% you will be asked to supply a shorter one by inserting either \authorrunning{} or
%% \titlerunning{} before \maketitle. Anyway, you can specify your own heads.
%%
%%
%% Note: In the following text body of your manuscript, please note several differences from
%%       other major journals:
%% (1) \subsection{Please Capitalize the First Letter of Each Notional Word in Subsection Title}
%% (2) Please Capitalize the First Letter of Each Notional Word in all tables' captions

%
%________________________________________________ sections below
%% Authors can give a citation as 'Michel et al. 1992'.
%% You may also use \cite, \citep and \citet for citation, and use Table~1 or Figure~1
%% and so forth. Using \ref and \label for cross-references of Tables/Figures
%% is a good way in adjusting/adding/removing text, tables or figures.
%

\section{Introduction}
Solar energetic particle (SEP) is one of the serious radiation hazard for the spacecraft and the astronauts in space. The relationship between SEP and solar activities is also a central topic in space physics and space weather. SEP events are usually classified to two types according {to} their different acceleration processes: impulsive and gradual events, thought to be produced by solar flare and coronal mass ejection (CME)-driven shock respectively \citep{Reames95a, Reames99a}. The gradual SEP events usually present the distinctive features, such as high peak flux intensity, high energy, long duration, and et al, comparing to impulsive events \citep[e.g.][]{Reames95a, Reames99a, Kahler96, Kahler05b}.
However, in many cases \citep[e.g.][]{Cane.etal03,Li.etal07b,Li.etal07a,Ding.etal16}, gradual and impulsive SEP components are mixed, which can not be distinctively classified to these two types.
{So the solar source of energetic particles in the large SEP events is still a popular issue. Some statistical results implied that the higher energetic particles (e.g. $>30$MeV) are dominantly accelerated by the concurrent solar flares, while the CME-driven shock is generally as an effective accelerator mainly for SEPs within lower energies }\citep[e.g.][]{Le.etal17,Le.Zhang17}.
{The spectra rigidity of GLE also revealed that the flare plays an important role in large gradual SEP event }\citep{Wu.Qin18}.
{Case study presented the evidence that the first arriving relativistic and non-relativistic protons and electrons are accelerated by the concurrent flare according to the timing analysis in an individual SEP event, and then these particles may be further accelerated by the associated CME-driven shock }\citep{Zhao.etal18}.

In general, a large SEP events, e.g. $I_p>10$pfu (pfu=$proton/cm^2~s~sr$) at $>10$MeV in GOES observations, are always almost associated with fast and wide CME eruptions, but inversely not all fast and wide CMEs can produce SEP events. So a number of mechanisms of CME generating SEPs were proposed, such as coronal waves, CME lateral expansion, CME-CME interaction, and so on \citep{Desai.Giacalone16, Lugaz.etal17}.

The intensity of large gradual SEP event is positively correlated with the speed of associated CME, but the scatter is very large \citep{Kahler96, Kahler.etal00}. This suggested that the number of ambient energetic particles may be another
%\del{deciding}
factor {determining the intensity of the associated SEP event} besides of associated CME-driven shock speed \citep[e.g][]{Kahler01,Kahler.Vourlidas14}. These seed particles may be from solar flares \citep{Mason.etal99, Mason.etal00} or from the preceding CMEs \citep{Gopalswamy.etal02, Gopalswamy.etal04, Li.etal12}.

CME {interaction} is a frequent phenomenon in solar corona and interplanetary (IP) space. Usually CME ``cannibalism'' or collision can happen in the process of two CMEs interaction \citep[e.g.][]{Gopalswamy.etal01, Temmer.etal14, Shanmugaraju.etal14}. \citet{Shen.etal12} presented a case of two CMEs colliding in IP space and revealed that these two magnetized plasmoids collided as if they were solid-like objects, with a likelihood of 73\% that the collision was super-elastic. In a study of the first ground level enhancement event (GLE) of solar cycle 24, which occurred on 17 May 2012, \citet{Shen.etal13} reported two CME erupting from a complicated active region separated by only $3$ minutes using the observations of STEREO and SOHO. Successive CMEs can also cause an extreme space weather storm in IP space via interaction and pre-conditioning of the interplanetary medium at the CMEs \citep{Liu.etal14}. \citet{Gopalswamy.etal02, Gopalswamy.etal03} suggested that CME interaction is an important aspect of SEP production, which can be as a good discriminator between SEP-poor and SEP-rich CMEs. However, \citet{Richardson.etal03} {argued that this interaction do not play a fundamental role in determining whether a wide and fast CME is associated with an SEP event.}

\citet{Gopalswamy.etal04} showed that there exists a strong correlation between high particle intensities and the presence of preceding CMEs with $24$ hours. And it is interpreted that seed particles may be trapped in the closed field lines of the preceding CMEs or associated turbulence so that they are subject to repeated acceleration by the shock driven by the second CME.
However, this time {interval between two CME eruptions} is too long to {make sure that} direct CME(shock)-CME interaction {is responsible for} the observed {large} SEP events,
because most large SEPs are believed to occur below $\sim10R_s$ ($R_s$ is solar radius) \citep{Kahler03}. Later, \citet{Li.Zank05a} suggested that two consecutive CMEs may provide a favorable environment for particle acceleration. Subsequently, \citet{Li.etal12} proposed the ``twin-CME'' scenario, where two CMEs erupt in sequence from the same or nearby active regions within a short period of time, the preceding CME or its shock can increase the turbulence level and/or seed population ahead of the main CME-driven shock. Thereby the enhanced turbulence level and seed population favor a more efficient particle acceleration at the main CME shock comparing to a single CME. \citet{Ding.etal13} extended the work of \citet{Li.etal12} and found that CMEs having a preceding CME with speed $>300$km/s within $9$ hours from the same active region have larger probability of leading to large SEP events than CMEs that do not have preceding CMEs. A subsequent case study showed that the SEP release time near the Sun is consistent with the time of the main CME leading edge overtaking the tailing edge of the preceding CME, as well as the radio enhancement \citep{Ding.etal14}.

Type II radio bursts have been often used as a diagnostic of the CME-driven shock in studying SEP events \citep[e.g.][]{Kahler82, Gopalswamy.etal05b, Cho.etal08}. Metric type II radio bursts are generated when the shock is close to the Sun (e.g. $\leq3R_s$) \citep{Gopalswamy.etal09a}. While many SEP events have metric type II bursts associated with them, the signature of metric type II bursts do not necessarily lead
to a large SEP event \citep{Kahler82}. \citet{Cliver.etal04} argued that the presence of the decameter-hectometric (DH) type II radio emissions may be used as a marker to distinguish between SEP-rich and SEP-poor metric type II radio bursts. Later, \citet{Gopalswamy.etal05b} found that CME tend to be more energetic if radio bursts appear from metric to DH wavelength. Usually, shock that survive beyond $3R_s$, indicated by the signature of type II radio emission from metric drifting to DH wavelength, are more stronger and broader \citep[e.g.][]{Cliver.etal04}.

  {CME interaction} can lead to radio enhancement following an IP type II burst when a fast CME overtaking a slow one, which may imply a strengthened shock \citep{Gopalswamy.etal01}(also see \citet{Shen.etal13, Ding.etal14, Temmer.etal14}). The result of \citet{Temmer.etal14} indicated that the interaction process is strongly position angle (PA) dependent in terms of timing as well as kinematical evolution, and the timing for the enhanced type II bursts may be related to shock streamer interaction.

In previous statistical works of CME interaction and its role on SEP production by, e.g., \citet{Gopalswamy.etal02, Gopalswamy.etal03, Gopalswamy.etal04, Ding.etal13},  the observations of CMEs and SEPs were all made only by spacecraft near the Earth. However, the CME projection effect and the longitude dependence of SEP flux detection are always inevitable, especially in the study of CME interaction. In this paper, we
make use of multiple spacecraft observations, and focus on the effect of {CME interactions} on the association with SEP events and radio enhancement by using SOHO and STEREO-A/B data.
STEREO A and B spacecraft are advancing ahead of or lagging Earth at $\sim1$~AU in the heliocentric orbits respectively, and separating slowly from the Earth by $\sim22^\circ~year^{-1}$. During the study period from January 2010 to August 2014, the separation between STEREO-A(B) and Earth increases from $\sim64^\circ(68^\circ$) to $\sim166^\circ(161^\circ)$.
Our paper is organized as follows: Section 2 presents the dataset; Section 3 shows our statistical results; and Section 4 contains the discussion and the conclusion.

\section{Dataset and analysis}
\label{sec.data}
From the online CME catalog CDAW (https://cdaw.gsfc.nasa.gov/CME\_list/), we identified 64 interacting CME pairs from January 2010 to August 2014 which satisfied the following criteria:
(1) the time interval between the main and preceding CMEs is less than 14 hours \citep{Ding.etal14b}.
(2) The angular width (WD) of both the main CME (fast one) and the preceding CME (pre-CME, slow one) are larger than $60$ degrees, ensuring that these two CMEs can be clearly identified simultaneously by SOHO/LASCO \citep{Brueckner.etal95} and STEREO/SECCHI \citep{Howard.etal08} images; (3) the main CME can overtake the preceding CME within the field of view (FOV) of both SOHO/LASCO C2-C3 and STEREO/SECCHI COR1-COR2 instruments (i.e. height~$\leq\sim30R_s$). This criterion removes false interaction due to projection effect from single view point; (4) the angular span area of the preceding CME needs to be intersected partly or fully by the main CME; to ensure the interaction.
We note that the dataset of this study is a subset of the twin-CME database introduced in \citet[e.g.][]{Ding.etal13}.

% Table list section -----------
\begin{landscape}
\begin{table}[!htbp]
\centering
\caption{The properties of two interacting CMEs (2010-2014).}
\label{table.1}
\linespread{1.1}
\setlength{\tabcolsep}{3pt}
%\tablewidth{-5 pt}
\scriptsize

\begin{tabular}{ccccccccccccccccccc}
\toprule
\multirow{3}{*}{No.} &  \multicolumn{6}{c}{main CME} & \multicolumn{6}{c}{preceding CME} & & & &  &  &  \\
\cline{2-7}
\cline{8-13}
  &\multirow{2}{*}{onset time} & CPA & WD & $v$& $m$  & $E_k$ & \multirow{2}{*}{onset time} & CPA & WD & $v$& $m$ & $E_k$
  & ${H_{int}}^a$ & type II$^b$ & type II  &$I_{ps}$ &$I_{pa}$ & $P_{pb}$\\
&&(deg)&(deg)&(km/s)&(g)&(erg)&&(deg)&(deg)&(km/s)&(g)&(erg)& (Rs) &&enhancement&\multicolumn{3}{c}{(1/$cm^2~s~sr$~MeV)}\\
(1)&(2)&(3)&(4)&(5)&(6)&(7)&(8)&(9)&(10)&(11)&(12)&(13)&(14)&(15)&(16)&(17)&(18)&(19)\\
\midrule
 1	 & 2010/02/13 23:18	 & 290	 &  63	 & 1005	 &  2.7e+15	 &  1.3e+31	 & 2010/02/13 19:54	 & 290	 & 126	 &  247	 &  4.5e+15	 &  1.4e+30	 &  3.6	 &   --	 & N	 &      --	 &      --	&      --	 \\
 2	 & 2010/03/17 12:30	 & 259	 &  66	 &  870	 &  8.5e+14	 &  3.2e+30	 & 2010/03/16 22:30	 & 229	 &  76	 &  113	 &  2.2e+15	 &  1.4e+29	 & 12.0	 &   --	 & N	 &      --	 &      --	&      --	 \\
 3	 & 2010/08/18 05:48	 & 255	 & 184	 & 1471	 &  1.1e+16	 &  1.2e+32	 & 2010/08/18 00:24	 & 298	 &  88	 &  403	 &  7.6e+15	 &  6.1e+30	 & 16.7	 &   DH	 & N$^d$	 &  0.0076	 &  0.0057	&  0.0013	 \\
 4	 & 2011/02/01 23:24	 & 276	 & 360	 &  437	 &  1.5e+15	 &  1.4e+30	 & 2011/02/01 20:00	 & 288	 &  77	 &   79	 &  3.1e+14	 &  9.6e+27	 &  4.5	 &   --	 & N	 &      --	 &      --	&      --	 \\
 5	 & 2011/02/24 07:48	 &  70	 & 158	 & 1186	 &  7.5e+15	 &  5.3e+31	 & 2011/02/24 02:36	 &  56	 &  71	 &  146	 &  2.8e+13	 &  3.0e+27	 &  7.0	 &    M	 & N	 &      --	 &      --	&      --	 \\
 6	 & 2011/04/17 01:25	 &  46	 &  75	 &  218	 &  5.6e+14	 &  1.3e+29	 & 2011/04/17 00:00	 & 359	 &  96	 &  179	 &  2.4e+15	 &  3.8e+29	 &  3.5	 &   --	 & N	 &      --	 &      --	&      --	 \\
 7	 & 2011/05/09 20:57	 &  55	 & 292	 & 1318	 &  1.0e+16	 &  8.8e+31	 & 2011/05/09 07:36	 &  58	 &  98	 &  132	 &  6.3e+15	 &  5.5e+29	 & 13.3	 &   DH	 & Y	 &      --	 &      --	&  0.0016	 \\
 8	 & 2011/05/12 13:25	 &  33	 &  95	 &  274	 &  5.4e+14	 &  2.0e+29	 & 2011/05/12 08:36	 & 130	 & 110	 &  166	 &  4.0e+15	 &  5.5e+29	 &  4.3	 &   --	 & N	 &      --	 &      --	&      --	 \\
 9	 & 2011/06/02 08:12	 &  99	 & 360	 &  976	 &  1.4e+15	 &  6.8e+30	 & 2011/06/02 07:24	 &  74	 &  61	 &  253	 &  1.4e+14	 &  4.4e+28	 &  3.5	 &   DH	 & N	 &      --	 &      --	&      --	 \\
10	 & 2011/09/06 23:05	 & 306	 & 360	 &  575	 &  1.5e+16	 &  2.5e+31	 & 2011/09/06 21:24	 & 289	 & 115	 &  499	 &  9.7e+15	 &  1.2e+31	     & 21.1	 & DH	 & Y	 &  0.0663	&      --	&  0.0048	 \\
11	 & 2011/10/04 13:25	 & 379	 & 360	 & 1101	 &  1.6e+16	 &  9.9e+31	 & 2011/10/04 12:12	 &  30	 & 124	 &  393	 &  6.2e+14	 &  4.8e+29	 &  4.8	 &   DH	 & Y	 &      --	&  0.0162	&  0.1155	 \\
12	 & 2011/10/14 12:24	 &  32	 & 241	 &  814	 &  7.9e+15	 &  2.6e+31	 & 2011/10/14 09:12	 &  40	 & 208	 &  454	 &  1.1e+16	 &  1.1e+31	 & 19.6	 &   DH	 & N	 &      --	&      --	&      --	 \\
13	 & 2011/10/20 03:36	 & 298	 & 193	 &  893	 &  4.6e+15	 &  1.8e+31	 & 2011/10/19 21:48	 & 302	 &  60	 &  239	 &  2.4e+15	 &  6.8e+29	 & 11.8	 &   --	 & N	 &      --	&      --	&      --	 \\
14	 & 2011/10/22 01:25	 & 313	 & 360	 &  593	 &  1.6e+16	 &  2.7e+31	 & 2011/10/21 17:48	 & 242	 &  96	 &  292	 &  1.1e+16	 &  4.7e+30	 & 27.1	 &   --	 & N	 &      --	&      --	&      --	 \\
15	 & 2011/11/07 23:48	 & 303	 & 109	 &  527	 &  3.2e+15	 &  4.4e+30	 & 2011/11/07 20:57	 & 280	 &  82	 &  305	 &  9.8e+14	 &  4.6e+29	 & 10.5	 &   --	 & N	 &      --	&      --	&      --	 \\
16	 & 2011/11/09 13:36	 &  62	 & 360	 &  907	 &  1.4e+16	 &  5.6e+31	 & 2011/11/09 08:36	 & 132	 & 147	 &  496	 &  1.3e+16	 &  1.6e+31	 & 29.2	 &   M,DH	 & Y£¿	 &      --	&      --	&  0.0027	 \\
17	 & 2011/11/26 07:12	 & 277	 & 360	 &  933	 &  1.2e+16	 &  5.2e+31	 & 2011/11/26 00:36	 & 250	 &  90	 &  292	 &  7.4e+15	 &  3.2e+30	 & 17.9	 &   DH	 & Y	 &  0.1017	&  0.0066	&  0.0020	 \\
18	 & 2011/12/09 12:12	 &  74	 & 117	 &  335	 &  9.7e+14	 &  5.4e+29	 & 2011/12/09 05:36	 & 163	 &  83	 &  155	 &  7.2e+14	 &  8.7e+28	 & 11.9	 &   --	 & N	 &      --	&      --	&      --	 \\
19	 & 2011/12/24 00:36	 &  56	 &  61	 &  475	 &  2.8e+15	 &  3.1e+30	 & 2011/12/23 18:36	 & 104	 &  74	 &  247	 &  2.2e+15	 &  6.9e+29	 & 14.1	 &   --	 & N	 &      --	&      --	&      --	 \\
20	 & 2012/01/12 08:24	 &  69	 & 360	 &  814	 &  1.0e+16	 &  3.4e+31	 & 2012/01/12 04:24	 &  79	 & 101	 &  280	 &  3.7e+15	 &  1.4e+30	 & 11.3	 &   DH	 & N	 &      --	&      --	&      --	 \\
21	 & 2012/01/19 14:36	 &  19	 & 360	 & 1120	 &  1.9e+16	 &  1.2e+32	 & 2012/01/19 09:48	 & 333	 & 111	 &  317	 &  9.9e+15	 &  5.0e+30	 & 13.7	 &   DH	 & N	 &      --	&      --	&  0.0197	 \\
22	 & 2012/01/23 04:00	 & 302	 & 360	 & 2175	 &  2.6e+16	 &  6.2e+32	 & 2012/01/23 03:12	 & 329	 & 221	 &  684	 &  5.3e+15	 &  1.2e+31	 &  6.6	 &   DH	 & Y	 & 10.3404	&  0.2319	&  0.4717	 \\
23	 & 2012/03/04 11:00	 &  52	 & 360	 & 1306	 &  7.9e+15	 &  6.8e+31	 & 2012/03/04 08:12	 &  61	 &  92	 &  207	 &       --	 &       --	 &  5.6	 &   M,DH	 & Y	 &  0.0026	&      --	&  0.2263	 \\
24	 & 2012/03/05 04:00	 & 391	 & 360	 & 1531	 &  1.4e+16	 &  1.6e+32	 & 2012/03/05 03:12	 &  29	 &  92	 &  594	 &  3.5e+15	 &  6.2e+30	 &  5.9	 &   DH	 & Y	 &  0.0085	&      --	&  0.1901	 \\
25	 & 2012/03/10 18:00	 & 276	 & 360	 & 1296	 &       --	 &       --	 & 2012/03/10 16:24	 & 292	 & 127	 &  423	 &  4.3e+15	 &  3.8e+30	 &  8.0	 &   M,DH	 & Y	 &      --	&      --	&      --	 \\
26	 & 2012/03/18 00:24	 & 303	 & 360	 & 1210	 &  1.1e+16	 &  7.9e+31	 & 2012/03/17 22:12	 & 348	 &  64	 &   66	 &  7.9e+14	 &  1.7e+28	 &  3.7	 & --	 & N	 &      --	&      --	&      --	 \\
27	 & 2012/03/28 01:36	 & 397	 & 360	 & 1033	 &  7.0e+15	 &  3.8e+31	 & 2012/03/28 00:48	 &  60	 & 126	 &  664	 &       --	 &       --	 & 12.6	 & --	 & N	 &      --	&      --	&      --	 \\
28	 & 2012/04/16 17:48	 &  62	 & 166	 & 1348	 &  7.3e+15	 &  6.7e+31	 & 2012/04/16 14:12	 &  55	 & 134	 &   89	 &  2.0e+15	 &  7.9e+28	 &  4.3	 &   --	 & N	 &      --	&      --	&      --	 \\
29	 & 2012/06/28 20:00	 &  42	 & 145	 & 1313	 &  7.3e+15	 &  6.3e+31	 & 2012/06/28 18:48	 & 127	 &  83	 &  343	 &  2.4e+15	 &  1.4e+30	 &  4.9	 &   DH	 & N	 &      --	&      --	&      --	 \\
30	 & 2012/07/17 13:48	 & 255	 & 176	 &  958	 &  1.7e+16	 &  7.8e+31	 & 2012/07/17 13:25	 & 153	 &  95	 &  292	 &  4.4e+15	 &  1.9e+30	 &  5.2	 &   M,DH	 & Y	 &  0.4160	&      --	&      --	 \\
31	 & 2012/08/04 13:36	 & 109	 & 360	 &  856	 &  1.3e+16	 &  4.7e+31	 & 2012/08/04 12:36	 & 124	 &  60	 &  187	 &  1.5e+15	 &  2.6e+29	 &  4.0	 &   --	 & N	 &      --	&      --	&      --	 \\
32	 & 2012/08/10 10:34	 & 230	 & 251	 &  464	 &  6.7e+15	 &  7.2e+30	 & 2012/08/10 00:44	 & 319	 &  92	 &  188	 &  4.7e+15	 &  8.2e+29	 & 19.9	 &   --	 & N	 &  0.0007	&  0.0045	&      --	 \\
33	 & 2012/08/18 00:48	 &  62	 & 145	 &  986	 &  3.6e+15	 &  1.8e+31	 & 2012/08/17 23:48	 & 338	 & 185	 &  463	 &  5.4e+15	 &  5.7e+30	 &  7.8	 &   --	 & N	 &      --	&      --	&      --	 \\
34	 & 2012/08/20 21:28	 &  68	 & 360	 &  521	 &  7.5e+15	 &  1.0e+31	 & 2012/08/20 19:36	 &  35	 &  75	 &  215	 &  1.5e+14	 &  3.5e+28	 &  5.8	 &   --	 & N	 &      --	&      --	&      --	 \\
35	 & 2012/08/26 11:12	 & 330	 & 143	 &  398	 &  1.2e+15	 &  9.6e+29	 & 2012/08/26 07:24	 &  25	 &  86	 &  208	 &  1.9e+15	 &  4.0e+29	 &  6.7	 &   --	 & N	 &      --	&      --	&      --	 \\
36	 & 2013/02/14 21:17	 & 274	 &  81	 &  690	 &  6.6e+15	 &  1.6e+31	 & 2013/02/14 19:24	 & 273	 &  60	 &  200	 &  2.0e+15	 &  3.9e+29	 &  5.6	 &   --	 & N	 &      --	&      --	&      --	 \\
37	 & 2013/03/05 03:48	 & 424	 & 360	 & 1316	 &  1.9e+16	 &  1.7e+32	 & 2013/03/05 00:36	 &  90	 &  81	 &   80	 &  2.4e+15	 &  7.8e+28	 &  3.9	 & M,DH	 & Y	 &  0.0024	& 11.0666	&  0.1183	 \\
38	 & 2013/03/24 15:48	 & 265	 &  73	 &  491	 &  8.0e+14	 &  9.7e+29	 & 2013/03/24 07:48	 & 335	 &  62	 &  180	 &  1.9e+15	 &  3.1e+29	 &  7.0	 &   --	 & N	 &      --	&      --	&      --	 \\
39	 & 2013/04/20 06:00	 & 283	 & 153	 &  741	 &  1.1e+16	 &  3.0e+31	 & 2013/04/20 03:48	 & 284	 &  73	 &  133	 &  1.4e+15	 &  1.3e+29	 &  5.0	 &   --	 & N	 &      --	&      --	&      --	 \\
40	 & 2013/04/26 22:00	 & 256	 &  67	 &  561	 &  2.8e+15	 &  4.4e+30	 & 2013/04/26 18:24	 & 200	 & 150	 &  271	 &  5.7e+15	 &  2.1e+30	 &  5.8	 &   --	 & N	 &      --	&      --	&      --	 \\
\bottomrule
\end{tabular}
\end{table}
\end{landscape}

\begin{landscape}
\begin{table}[!htbp]
{\bf Table~\ref{table.1}} (Continued)  The properties of two interacting CMEs (2010-2014).
\label{table.2}
\linespread{1.1}
\setlength{\tabcolsep}{3pt}
%\tablewidth{-5 pt}
\scriptsize
\centering
\begin{tabular}{ccccccccccccccccccc}
\toprule
\multirow{3}{*}{No.} &  \multicolumn{6}{c}{main CME} & \multicolumn{6}{c}{preceding CME} & & & &  &  &  \\
\cline{2-7}
\cline{8-13}
  &\multirow{2}{*}{onset time} & CPA & WD & $v$& $m$  & $E_k$ & \multirow{2}{*}{onset time} & CPA & WD & $v$& $m$ & $E_k$
  & ${H_{int}}^a$ & type II$^b$ & type II &$I_{ps}$ &$I_{pa}$ & $P_{pb}$\\
&&(deg)&(deg)&(km/s)&(g)&(erg)&&(deg)&(deg)&(km/s)&(g)&(erg)& (Rs) &&enhancement&\multicolumn{3}{c}{(1/$cm^2~s~sr$~MeV)}\\
(1)&(2)&(3)&(4)&(5)&(6)&(7)&(8)&(9)&(10)&(11)&(12)&(13)&(14)&(15)&(16)&(17)&(18)&(19)\\
\midrule
41	 & 2013/05/17 09:12	 &  74	 & 360	 & 1345	 &  5.8e+15	 &  5.2e+31	 & 2013/05/17 04:49	 &  78	 &  86	 &  107	 &  1.5e+15	 &  8.4e+28	 &  5.0	 &   M	 & N	 &      --	&      --	&      --	 \\
42	 & 2013/05/22 13:25	 & 251	 & 360	 & 1466	 &  3.3e+16	 &  3.5e+32	 & 2013/05/22 08:48	 & 270	 & 210	 &  687	 &  3.1e+16	 &  7.3e+31	 &  6.4$^c$	 & M,DH	 & Y	 &  8.2122	 &  0.0355	&      --	 \\
43	 & 2013/07/01 20:24	 & 123	 & 208	 &  819	 &  1.3e+16	 &  4.5e+31	 & 2013/07/01 08:12	 &   7	 &  98	 &   71	 &  3.1e+15	 &  7.7e+28	 &  7.3	 &   --	 & N	 &      --	&  0.0018	&  0.0056	 \\
44	 & 2013/07/06 19:36	 & 147	 & 123	 &  380	 &  2.7e+15	 &  1.9e+30	 & 2013/07/06 12:36	 & 101	 & 236	 &  127	 &  9.0e+14	 &  7.2e+28	 & 16.1	 &   --	 & N	 &      --	&      --	&      --	 \\
45	 & 2013/07/18 20:24	 & 102	 &  63	 &  939	 &  4.1e+15	 &  1.8e+31	 & 2013/07/18 18:24	 & 102	 & 135	 &  458	 &  4.9e+15	 &  5.2e+30	 & 11.0	 &   --	 & N	 &      --	&      --	&      --	 \\
46	 & 2013/07/29 13:25	 &  42	 & 145	 &  542	 &  2.9e+15	 &  4.2e+30	 & 2013/07/29 10:36	 & 155	 & 124	 &  226	 &  1.7e+15	 &  4.4e+29	 &  7.3	 &   --	 & N	 &      --	&      --	&      --	 \\
47	 & 2013/10/11 07:24	 &  88	 & 360	 & 1200	 &  1.0e+16	 &  7.3e+31	 & 2013/10/11 01:36	 &  21	 & 108	 &  269	 &  4.9e+15	 &  1.8e+30	 & 11.5	 &   M,DH	 & Y	 &  0.0011	&  2.1588	&  0.2092	 \\
48	 & 2013/10/24 01:25	 & 122	 & 360	 &  399	 &  1.9e+15	 &  1.5e+30	 & 2013/10/23 23:12	 &  93	 & 110	 &  162	 &  1.3e+15	 &  1.7e+29	 &  4.8	 &   M	 & N	 &      --	&      --	&      --	 \\
49	 & 2013/10/28 04:48	 & 315	 & 156	 & 1201	 &  6.3e+15	 &  4.5e+31	 & 2013/10/28 02:24	 & 265	 & 360	 &  695	 &  8.6e+15	 &  2.1e+31	 & 26.1	 &   DH	 & Y	 &  0.0143	&  0.0031	&      --	 \\
50	 & 2013/12/26 03:24	 & 130	 & 360	 & 1336	 &  5.6e+15	 &  5.0e+31	 & 2013/12/26 03:12	 & 122	 & 171	 & 1022	 &  9.4e+15	 &  4.9e+31	 &  8.7	 &   DH	 & Y	 &  0.0107	&  0.2317	&  0.1907	 \\
51	 & 2014/01/06 00:36	 & 276	 & 129	 &  574	 &  1.9e+15	 &  3.2e+30	 & 2014/01/05 19:00	 & 301	 & 101	 &  269	 &  2.5e+15	 &  9.1e+29	 & 17.3	 &   --	 & N	 &      --	&      --	&      --	 \\
52	 & 2014/01/16 23:36	 & 114	 & 197	 &  666	 &  1.1e+16	 &  2.5e+31	 & 2014/01/16 21:38	 &  95	 & 113	 &  438	 &  2.1e+15	 &  2.0e+30	 & 11.6	 &   --	 & N	 &      --	&      --	&      --	 \\
53	 & 2014/01/25 16:48	 & 152	 & 158	 &  783	 &  2.5e+15	 &  7.8e+30	 & 2014/01/25 14:48	 & 127	 &  80	 &  299	 &  4.8e+14	 &  2.1e+29	 &  7.3	 &   --	 & N	 &      --	&      --	&      --	 \\
54	 & 2014/01/30 16:24	 & 120	 & 360	 & 1087	 &  1.0e+16	 &  6.0e+31	 & 2014/01/30 15:48	 & 109	 &  62	 &  780	 &  3.0e+15	 &  9.3e+30	 &  7.1	 &    M	 & N	 &  0.0008	&      --	&      --	 \\
55	 & 2014/01/30 08:24	 & 117	 & 360	 &  458	 &  4.5e+15	 &  4.7e+30	 & 2014/01/30 06:12	 &  64	 &  75	 &  116	 &  1.1e+14	 &  7.6e+27	 &  4.1	 &   --	 & N	 &      --	&      --	&      --	 \\
56	 & 2014/02/21 16:00	 &  86	 & 360	 & 1252	 &  7.2e+15	 &  5.6e+31	 & 2014/02/21 12:12	 & 184	 &  62	 &  341	 &  8.4e+14	 &  4.9e+29	 & 11.3	 &   DH	 & N	 &      --	&  0.0038	&  0.0472	 \\
57	 & 2014/03/14 03:24	 &  58	 &  66	 &  314	 &  1.7e+15	 &  8.5e+29	 & 2014/03/14 01:36	 & 133	 &  91	 &  110	 &  1.4e+15	 &  8.4e+28	 &  4.2	 &   --	 & N	 &      --	&      --	&      --	 \\
58	 & 2014/03/22 10:00	 & 323	 & 169	 &  756	 &  5.9e+15	 &  1.7e+31	 & 2014/03/22 06:48	 & 257	 & 168	 &  340	 &  7.8e+15	 &  4.5e+30	 & 10.0	 &   --	 & N	 &      --	&      --	&      --	 \\
59	 & 2014/04/05 00:12	 & 155	 & 149	 &  585	 &  4.0e+15	 &  6.8e+30	 & 2014/04/04 21:12	 &  58	 &  73	 &  359	 &  5.9e+14	 &  3.8e+29	 &  8.4	 &   --	 & N	 &  0.0025	&      --	&      --	 \\
60	 & 2014/05/09 02:48	 & 276	 & 360	 & 1099	 &  1.3e+16	 &  7.7e+31	 & 2014/05/09 01:25	 & 236	 &  92	 &  161	 &  1.8e+15	 &  2.4e+29	 &  6.6	 &   DH	 & Y	 &  0.0020	&      --	&      --	 \\
61	 & 2014/07/20 03:12	 &  72	 & 135	 &  417	 &  5.2e+15	 &  4.6e+30	 & 2014/07/19 20:12	 &  62	 &  76	 &  135	 &  1.8e+15	 &  1.7e+29	 & 11.0	 &   --	 & N	 &      --	&      --	&      --	 \\
62	 & 2014/07/30 16:12	 & 139	 & 143	 &  638	 &  4.4e+15	 &  8.9e+30	 & 2014/07/30 13:36	 &  41	 & 124	 &  274	 &  4.0e+15	 &  1.5e+30	 &  7.0	 &   --	 & N	 &      --	&      --	&      --	 \\
63	 & 2014/08/25 15:36	 & 272	 & 360	 &  555	 &       --	 &       --	 & 2014/08/25 13:48	 & 252	 &  75	 &  315	 &  3.9e+14	 &  1.9e+29	 &  8.4	 &   M,DH	 & N	 &  0.0053	&      --	&      --	 \\
64	 & 2014/08/28 17:24	 & 415	 & 360	 &  766	 &  1.0e+16	 &  2.9e+31	 & 2014/08/28 11:48	 &  13	 &  93	 &  190	 &  3.7e+15	 &  6.7e+29	 & 10.4	 &   DH	 & Y	 &  0.0005	&  0.1472	&  0.1781	 \\
\bottomrule
\end{tabular}
\begin{itemize}
\scriptsize
  \item[$a$] The intersection height of two CMEs leading-edge trajectories. The quadratic fitting is used for relationship between CME height and time.
  \item[$b$] Type II radio bursts associated with main CME. M -- metric type II radio burst observed by ground stations, such as Learmonth, Palehua, San-vito, Sagamore-hill; DH -- DH type II radio burst observed by Wind/WAVES, or STEREO/SWAVES. `--' denote no type II radio burst.
  \item[$c$] This value is cited from \citep{Ding.etal14}
\end{itemize}
\end{table}
\end{landscape}

The properties of main CMEs and preceding CMEs are listed in Table~\ref{table.1}, which are taken from the CDAW online database. Column 1 is the event number; column 2 is the onset time (first appearance in LASCO/C2) of the main(preceding) CME; column 3 is the central position angle (CPA); column 4 is the angular width (WD); column 5 is the CME speed ($v$) projected to the sky plane from the view at the Earth; column 6-7 are the mass ($m$) and kinetic energy ($E_k$) respectively.
Column 8 to 13 show the similar parameters for the preceding CME.
By using a quadratic polynomial fit to the height-time measurements of the CMEs at CDAW, the intersection heights of leading-edge trajectories of the main and the preceding CME are obtained, which
are listed in column 14 of Table~\ref{table.1} (and labeled as $H_{int}$).

   \begin{figure}[htb]
   \centering
   \includegraphics[width=0.5\textwidth, angle=0]{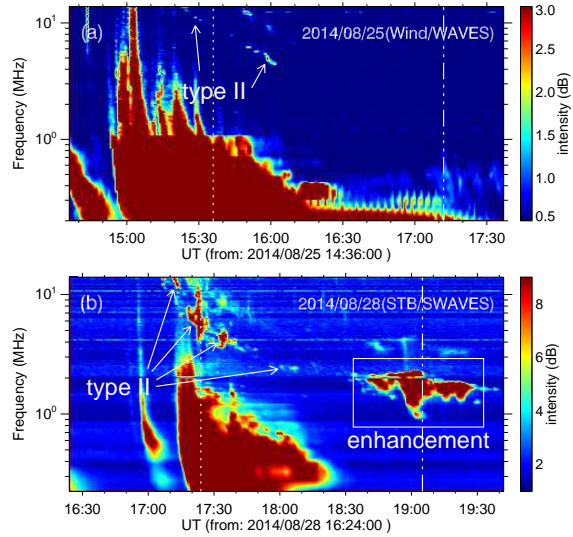}
   \caption{A couple of examples displaying the type II radio bursts without (a) or with (b) enhancement during CME interactions. The dash line denotes the onset time of the associated CME in LASCO C2 field of view (FOV), and the dot-dash line denotes the intersection time of leading-edge trajectories of the main CME crossing the pre-CME.}
   \label{Fig1.exam}
   \end{figure}

Type II radio bursts yield information on the formation and the propagation of the CME associated shock. For example, Figure~\ref{Fig1.exam} shows the dynamic spectra of the events on 25 August 2014 and 28 August 2014 within decameter-hectometric (DH) wavelength, detected by Wind/WAVES and STB/SWAVES respectively. The feature of frequency shifting denoted by the arrows, are distinctly clear. Note,however the type II radio intensities in panel (a) are weaker and (b) are stronger. And a continuum-like enhancement of decametric to hectometric type II radio emission presents in the low frequency for the event in panel (b) (denoted by box), which may be interpreted as observational signature of the transit of the shock front of the fast CME through the core of a slow CME as the consequence of {CME interactions} (also see \citet{Gopalswamy.etal01}).
This unusual enhanced spectral continuum in DH wavelength is the main subject of this paper and is defined as the enhanced type II radio bursts or type II radio enhancement. So in Figure~\ref{Fig1.exam}, the event in panel (a) is associated with a type II radio burst but no enhancement, while the event in panel (b) is associated with an enhanced type II radio burst. To ensure that the type II radio enhancement indeed corresponds to the interaction, we compare the time of trajectory intersection to the time of enhancement. Only when the intersection time (e.g. indicated by the dot-dash lines in Figure~\ref{Fig1.exam}) is behind the start time of radio burst enhancement which means the enhancement is occurred during the main CME transiting through the body of preceding CME, radio enhancement is identified to be associated with the CME interaction. Similar cases can also be seen from \citet{Gopalswamy.etal01, Ding.etal14}.

Type II radio bursts associated with the main CMEs are listed in column 15. `M' denotes the metric type II radio burst detected by the ground stations, such as Learmonth, Palehua, San-vito, Sagamore-hill {(https://www.ngdc.noaa.gov/stp/space-weather/solar-data/solar-features/solar-radio)}. `DH' denotes the DH type II radio burst observed by Wind/WAVES {(ftp://cdaweb.gsfc.nasa.gov/pub/data/wind/waves)} \citep{Bougeret.etal95} and/or STEREO-A(B)/SWAVES {(ftp://stereoftp.nascom.nasa.gov/pub/ins\_data/swaves)} \citep{Bougeret.etal08}. Column 16 displays whether the type II radio bursts enhancement is due to because of the interaction of two CMEs (Y-yes or N-no).

   \begin{figure}[htb]
   \centering
   \includegraphics[width=0.7\textwidth, angle=0]{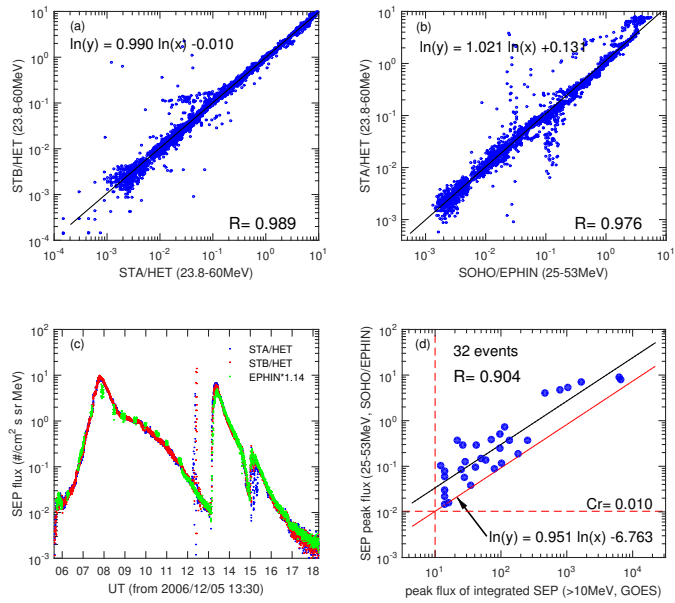}
   \caption{Comparison of the differential solar energetic proton intensities in STEREO/HET and SOHO/EPHIN using observations during December 2006, and the peak intensity of SEP events  from 2010 to 2014 between SOHO/EPHIN and GOES. (a) STB/HET vs. STA/HET in 23.8-60 MeV; (b) STA/HET 23.8-60 MeV vs. SOHO/EPHIN 25-53 MeV; (c) energetic proton flux time profiles in STA 23.8-60 MeV (blue), STB 23.8-60 MeV (red), and SOHO 25-53 MeV with multiplying 1.14 (green); (d) peak flux of proton intensities in SOHO/EPHIN 25-53 MeV vs. that in GOES $>10$MeV for large SEP events. The black lines are the linear fits to the data; the red solid line indicates the lower limit of data, which is shifted from the linear fit with the same slope.}
   \label{Fig2.cal}
   \end{figure}

%Before the STEREO era, only the instruments near the Earth can be used to identify the SEP events, e.g. GOES/EPS, ACE/SIS.
During the STEREO era, SEP observations were made at multiple locations at 1 AU and observations of the far side of the Sun were available, allowing the far side sources of SEPs to be identified \citep{Lario.etal13, Richardson.etal14}.
This leads to a great advantage of single-point observations. For single point observation, the quality of magnetic connection between the source location and the spacecraft often decides the flux level of an event and therefore if the event can be classified as SEPs. This is particularly true for backside events and small events \citep{Reames.etal96}.
For multi-spacecraft events, we can construct a better criteria to identify cases as large SEP events. In this work, we use the energetic particle observations made by the HET instrument onboard STEREO-A/B(STA/STB) \citep{VonRosenvinge.etal08} and the EPHIN instrument onboard SOHO \citep{Mueller-Mellin.etal95}. For STEREO, we focus on the energy range $23.8$-$60$ MeV, defined by a combination of HET energy channels to enhance the counting statistics. For SOHO, we use the energy range $25$-$53$ MeV of the EPHIN instrument, which matches 23.8-60MeV on STEREO/HET. The intercalibration between the various instruments used in this study can be checked over a wide dynamic range during the events in December 2006, when the STEREO spacecraft were still close to the Earth \citep[e.g.][]{Lario.etal13, Richardson.etal14}. Figure~\ref{Fig2.cal} shows the results of the comparison between different instruments in the energy range above. The black solid lines are the linear fits to the data. Figure~\ref{Fig2.cal}(a) displays five minutes average proton intensity from STA/HET plotted against STB/HET at the same energy range. The energetic proton intensities at both spacecraft are highly correlated (R=0.989), with the similar intensities (i.e. $I(B)=0.990I(A)^{0.990}$), which suggests that the observed intensities by STA and STB are comparable. Figure~\ref{Fig2.cal}(b) shows the 23.8-60MeV proton intensities from STA/HET plotted versus the 25-53MeV proton intensities from SOHO/EPHIN. The STA/HET intensity is correlated with (R=0.976) but $\sim1.14$ times higher than the SOHO/EPHIN intensity in a slightly narrow energy range. So in this study, the SOHO/EPHIN intensity is multiplied by a factor of $1.14$ to be compared with the STEREO/HET intensity, thus allowing us to compare intensities of SEP events measured by the selected instruments onboard different spacecraft. Figure~\ref{Fig2.cal}(c) presents the proton flux of SOHO/EPHIN (*1.14) and STEREO/HET from  2006 December 5 to 18, which shows a good agreement between
each other in both the ascending phases and the decay phases of the SEP events.

The integrated proton flux detected by GOES instrument are usually used to define a large SEP event (e.g. the peak flux $\geq10$pfu at $>10$MeV energy channel). To obtain the intensity threshold of identifying the large SEP events from STEREO/HET and SOHO/EPHIN at $\sim25-\sim60$MeV, we compare the SOHO/EPHIN 25-53MeV and GOES $>10$MeV peak intensities of $32$ large SEP events from 2010 to 2014, shown with a good correlation (R=0.904) in the panel (d) because of the close site of SOHO and GOES near the Earth. We then shift the linear fit line down to the data lower limit, denoted by the red solid line, given by an empirical formula $y=0.00114x^{0.951}$. Using this relationship, the criterion of large SEP event is obtained to be  $0.01$ ($1/cm^{2}~s~sr$~MeV) for SOHO/EPHIN at 25-53MeV.
We note that a few small events under the criterion of GOES may become SEP events using this threshold due to using the lower limit in Figure~\ref{Fig2.cal}(d).
 %meanwhile, the small events are defined below this value oppositely.
With the multiplicative factor 1.14 to SOHO/EPHIN observations, this value is set to 0.0114 ($1/cm^{2}~s~sr$~MeV) hereafter. The SEP peak intensities measured by different spacecraft are listed in column 17-19 of Table~\ref{table.1} respectively: $I_{ps}$ -- SOHO/EPHIN*1.14, $I_{pa}$ -- STA/HET, $I_{pb}$ -- STB/HET. Symbol `--' denotes that there was no detectable SEP event.

\section{Results}
\label{sec.res}

\subsection{Type II radio bursts associated with SEP events}
\label{sec.hist_rd}
Type II radio bursts are generally used to indicate whether a shock is formed during the CME eruption. When the shock propagates from low corona to high corona, the frequency of type II radio emission, decided by the ambient plasma density, decreases from metric wavelength to DH wavelength since the coronal plasma density decreases. The fast CME overtaking the preceding CME is usually accompanied by the type II radio enhancement \citep[e.g.][]{Gopalswamy.etal01,Ding.etal14}.
But not all cases of CME interaction lead to type II radio burst or enhanced type-II radio bursts, because radio emission or its enhancement also depends on the properties of CME pair, such as CME speed, kinetic energy, spatial relation, and so on.
Here, we classified all the 64 CME pairs in this study into two groups: (1) with enhanced type II radio burst (19 events: en-type-II events); (2) with type-II radio burst but no enhancement or without any type II radio burst (45 events: noen-type-II events).

For an individual SEP event, the detected peak intensity by a given spacecraft may be significantly effected by the longitudinal separation between the solar source location and the foot-points of magnetic field lines connecting the spacecraft to the Sun, or the longitude of the spacecraft relative to the location of the solar event. We therefore selected the maximum value among the peak intensities observed by three spacecraft (SOHO and STEREO-A/B) as the nominal intensity of this event (labeled by `$I_p$' hereafter). Comparing to single spacecraft observations, our method will decrease the chance of a large SEP event to be missed or underestimated due to poor magnetic connection.

   \begin{figure}[htb]
   \centering
   \includegraphics[width=0.8\textwidth, angle=0]{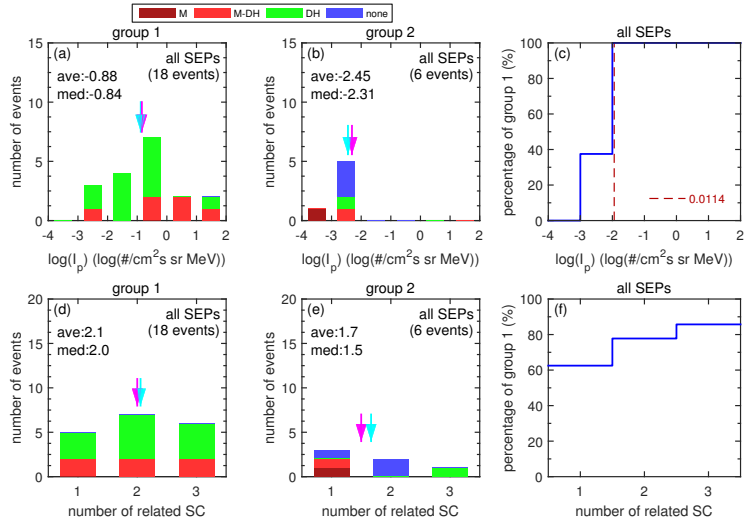}
   \caption{Histograms of the intensities ($I_p$) of SEP events associated with the two CMEs interactions are shown in panel (a,b). Histograms of the number of the different spacecraft that have detected the same SEP event are shown in panel (d,e). In each bar, dull red, red, green colors indicate the events that associated with type II radio bursts emitted in metric (M) wavelength only, metric and DH (M-DH), DH wavelength respectively; and blue denote the events with no detectable type II radio emission. Panel (c,f) show the percentage of the events with enhanced type II radio bursts in each statistical bin corresponding to left panels. The cyan and magenta arrows indicate the average and median values of the data respectively.}
   \label{Fig3.hist_I}
   \end{figure}

Among 24 SEP events listed in Table~\ref{table.1}, 18 events (including all 15 large SEP events) are associated with CME pairs assigned to group 1, and 6 events (only small SEP events) to group 2.  The histograms of the SEP log peak intensities for two groups are shown in Figure~\ref{Fig3.hist_I}(a-b, top). The colors of dark red, red, green, and blue indicate events that are associated with type II radio bursts in metric (M), metric-DH (M-DH), DH wavelength, and no type II radio signatures, respectively. The log $I_p$ of SEP events in group 1 varies from -2.80 to 1.04 (average -0.88, median -0.84), while log $I_p$ of SEP events in group 2 varies from -3.08 to -2.12 (average -2.45, median -2.31).
 It's clear that the intensity of SEP event with the presence of enhanced type II radio burst is more larger than that with the absence of type II radio emission or its enhancement.
\citet{Gopalswamy.etal05b} examined the role of the metric radio bursts and DH radio bursts in large SEP events. They found that CMEs tend to be more energetic if radio bursts appear in all three wavelength ranges (i.e. from m-to-km). Since the plasma frequency at $\sim3R_{s}$ is $\sim14$MHz, \citet{Cliver.etal04} suggested that the shocks that survive beyond $3R_{s}$ are more stronger and broader and therefore accelerate particles to high energies. From panel (a) and (b), we see clearly that the majority of SEP events associated with metric type II radio bursts (M and M-DH) in group 1 are more intense. This is not true for events in group 2. It's easily understood that the shock marked by metric type II radio bursts as a stronger accelerator can produce a larger SEP event if enough seed particles from the core of preceding CME are overtaken by the shock marked by the enhancement of type II radio bursts.
The panel (c) shows the percentage of the en-type-II SEP events in each $I_p$ bin, which indicates that all large SEP events (i.e. $I_p\geq~0.0114$) are all from group 1.
Here, one may ask which factor is correlates the most significantly with SEP intensity. The presence or absence of type II emissions itself or the presence or absence of enhancements of type II radio burst? From Table~\ref{table.1} and Figure~\ref{Fig3.hist_I}, we see that: $18/19$(95\%) of CME pairs with the presence of radio enhancements generate SEP events. In comparison, only $6/45$(13\%) of CME pairs with the absence of type II radio emissions or enhancements can generate SEP events. By comparison, $21/29$(72\%) of CME pairs with the presence of type II can lead to SEP events, and $3/35$(9\%) without can lead to SEP events. If only considering the large SEP events, $15/19$(79\%) CME pairs with enhanced type II can produce large SEP events, while only $15/29$(52\%) CME pairs with type II can produce large SEP events. Clearly, the presence of enhancements of type II radio bursts correlates better with the occurrence of SEP and its size, comparing to the presence of type II radio emission alone.
So it's suggested that, for interacting CME pairs, the majority of SEP events are almost generated by fast CMEs overtaking preceding CMEs and lead to type II radio enhancement, especially for large SEP events. The signature of type II radio burst enhancement during CME interactions seems to be a good discriminator between large and small or none SEP event producers.

We also examine the number of spacecraft that can observe the SEP flux increasing from the background at the energy range of $\sim25-\sim60$MeV, shown in Figure~\ref{Fig3.hist_I}(d-f, bottom). The number of SEP-observed spacecraft may be used roughly to indicate the longitudinal spreading of energetic particles and/or the shock strength.
From panel (f), the percentage of events accompanied by enhanced type II radio bursts rises from around 63\% for one-spacecraft events to around 78\% for two-spacecraft events, and around 86\% for three-spacecraft events. This could be a selection effect in that the shocks are more intense when type II radio enhancements are present {during} the {CME interactions}.

\subsection{Properties of en-type-II and noen-type-II interacting CMEs}
\label{sec.rden}
Since type-II radio burst enhancement can serve as a distinct signature of large SEP event occurrence during the interaction of CME pairs,
what properties of the main and preceding CME can be identified as key conditions resulting radio enhancement?

   \begin{figure}[htb]
   \centering
   \includegraphics[width=0.8\textwidth, angle=0]{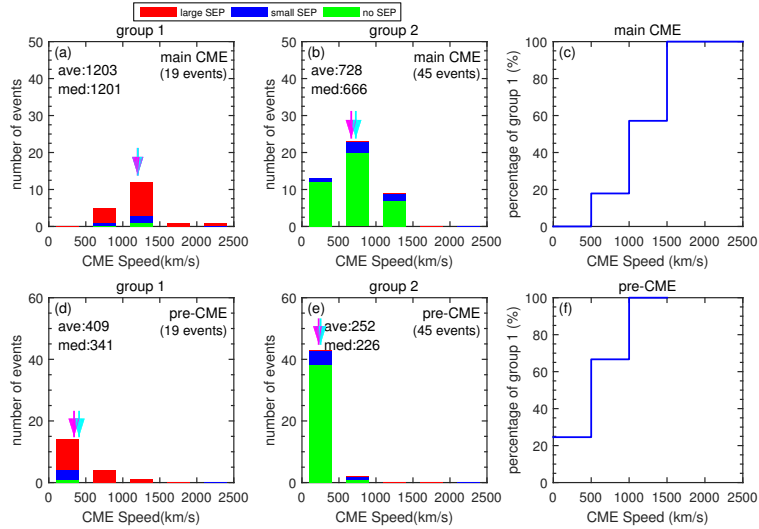}
   \caption{Histograms of the speed of main CMEs (panel (a,b)) and preceding CMEs (panel (d,e)). The cyan and magenta arrows indicate the average and median values respectively. The red, blue and green colors denote the large, small, and none SEP events respectively. Panel (c,f) present the percentage of the events that have enhanced type II radio bursts associated in each speed bin. }
   \label{Fig4.hist_v}
   \end{figure}

Figure~\ref{Fig4.hist_v} shows the speed distributions of the main CMEs (top) and the preceding CMEs (bottom) for en-type-II events (group 1) and noen-type-II events  events (group 2).
The speed of the main CMEs vary from 575km/s to 2175km/s in group 1, and 218km/s to 1471km/s in group 2. The difference of the average(median) speed of main CME between two groups is clearly distinct, i.e. 1203(1201)km/s and 728(652)km/s respectively.
The speed of the main CME associated with radio enhancement is generally larger than that with no radio emission or no enhancement.
However, the speed scattering is very large, and the speed of some CMEs in group~1 can be as low as $<900$km/s. Panel(c) presents the percentage of the en-type-II events in each speed bin.
For the main CMEs, the percentage of group 1 rises from only around 18\% for $v<1000km/s$ to around 57\% for $v<1500km/s$, and to 100\% for $v\geq1500km/s$,
 which shows
 that the probability of type II radio burst enhancement increases rapidly with the main CME speed. For the preceding CMEs, shown in panel (d-f), the average (median) speed of group 1 is also larger than that of group 2 (409(341) vs. 252(226)km/s).
The percentage of events in group 1 also shows the positive correlation with the speed of the preceding CMEs. It suggests that when two faster CMEs interact, the presence of the enhancement of type II radio bursts happens more likely.

The colors bars of red, blue, and green in Figure~\ref{Fig4.hist_v} denote the interacting CME pairs associated with large SEP events ($I_p\geq0.0114/cm^2~s~sr$~MeV), small SEP events ($I_p<0.0114/cm^2~s~sr$~MeV), and none SEP events. In group 1, about 79\%(15/19) pairs generate large SEP events, and about 21\%(4/19) pairs generate small SEP events, while in group 2, none of them produce large SEP event and only about $13$\%(6/45) pairs lead to small SEP events. The enhancement of type II radio burst hereby seems to be a good discriminator between SEP-rich (or large-SEP-rich) and SEP-poor interacting CME pairs.
%It's interesting to see that the speeds of main CMEs that produce large and small SEP events vary in a very large extent, e.g. from $\sim$575km/s to 2175km/s (in Group I) and from $\sim$555km/s to 1087km/s respectively.
%As shown here, especially from some main CMEs with very low speed, it seems to suggest that particle acceleration of interacting CMEs may be more efficiently than singles.
%This of course, is consistent with the proposal of the ``twin-CME'' scenario by \citet{Li.etal12} that the high-efficient production of large SEP event does not need a high speed main CME shock if there is enhanced turbulence and seed population level ahead of them, which is presumably provided by the preceding CME.

   \begin{figure}[htb]
   \centering
   \includegraphics[width=0.8\textwidth, angle=0]{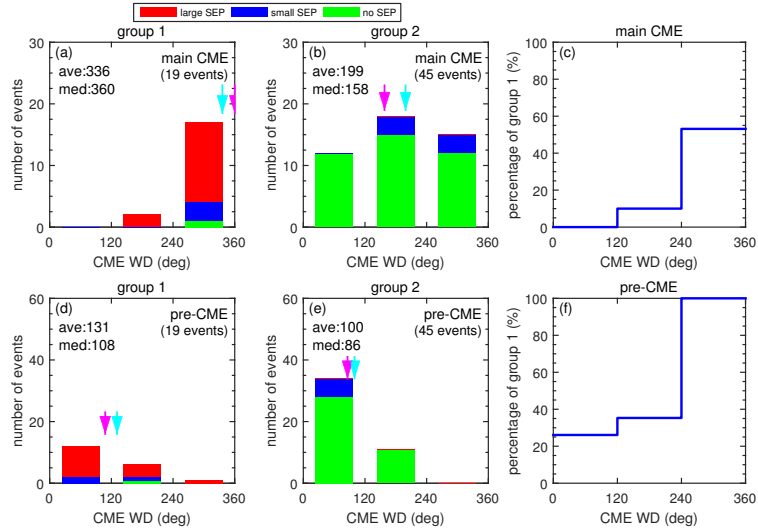}
   \caption{Same as Fig.~\ref{Fig4.hist_v}, but for the angular width (WD) of CME pairs. }
   \label{Fig5.hist_wd}
   \end{figure}

The CME angular width (WD) in the field of view (FOV) of SOHO/LASCO is shown in Figure~\ref{Fig5.hist_wd}. For the main CME (top panels), the typical WD of en-type-II events (average $336^\circ$, median $360^\circ$) is much larger than that of noen-type-II events (average $199^\circ$, median $158^\circ$). A great majority of en-type-II main CMEs (84\%, 16/19) are halo CMEs, comparing to the noen-type-II main CMEs (29\%, 13/45). As  shown in panel (c), the percentage of group 1 in the bin of large WD is distinctly higher than that of small WD.
The percentage of CME pairs associated with enhanced type II radio bursts rises from only about 10\% for the WD of main CME below $240^\circ$, to about 53\% for WD great than $240^\circ$, which may be attributed to the fact of high proportion of halo CMEs.
The preceding CMEs in group 1 have the similar distribution as that in group 2 (bottom panels): the number of events in each bin seems to decrease with increasing WD.
The en-type-II pre-CMEs are slight wider than the noen-type-II ones (average: $131^\circ$ vs. $100^\circ$, median: $108^\circ$ vs. $86^\circ$).
Here we must note that the angular width from CDAW is only the projected value measured in the plane-of-sky, which may be significantly different from the deprojected value \citep{Shen.etal13b}. In this study, we only compare the relative size for CME width in different groups, and do not use the absolute WD value.

   \begin{figure}[htb]
   \centering
   \includegraphics[width=0.8\textwidth, angle=0]{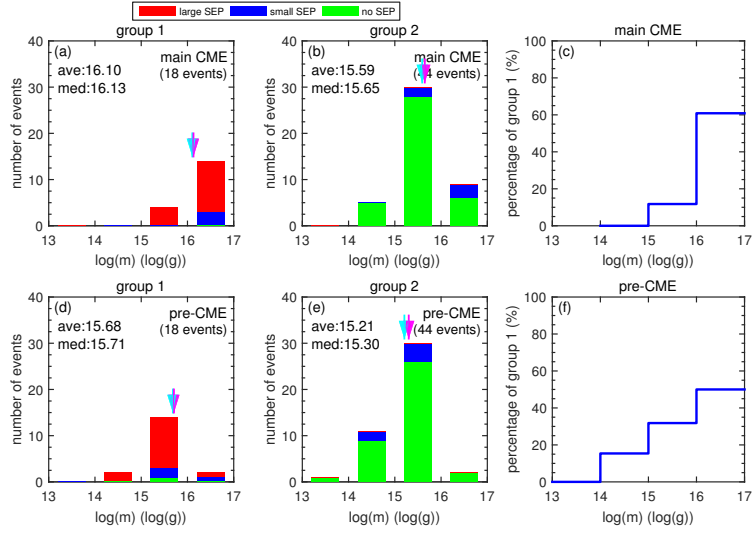}
   \caption{Same as Fig.~\ref{Fig4.hist_v}, but for the mass ($m$) of CME pairs. }
   \label{Fig6.hist_m}
   \end{figure}

   \begin{figure}[htb]
   \centering
   \includegraphics[width=0.8\textwidth, angle=0]{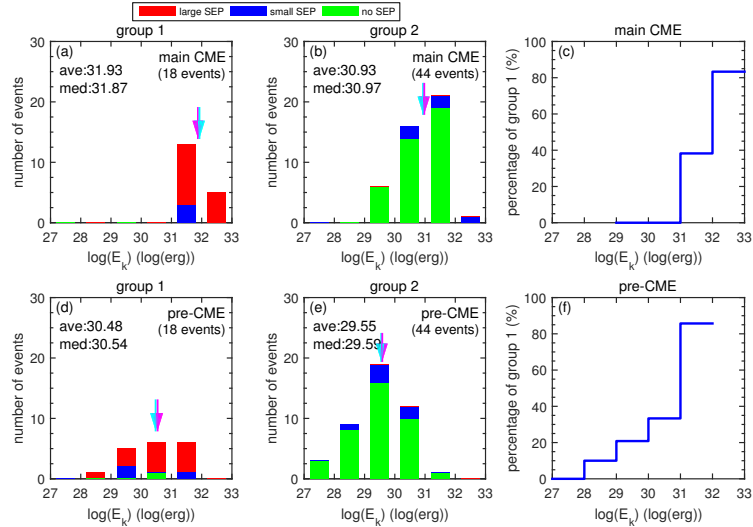}
   \caption{Same as Fig.~\ref{Fig4.hist_v}, but for the kinetic energy ($E_k$) of CME pairs. }
   \label{Fig7.hist_e}
   \end{figure}

The mass and kinetic energy of interacting CMEs may also be important factors for type II radio enhancement and/or the occurrence of SEP events. The statistical results are shown in Figure~\ref{Fig6.hist_m} and \ref{Fig7.hist_e}.

The log mass of main CMEs in group 1 (average 16.10, median 16.13) is typically higher than that in group 2 (15.59, 15.65), shown in Figure~\ref{Fig6.hist_m}(a,b)). From the percentage distribution of the radio-enhanced events along the log mass, shown in Figure~\ref{Fig6.hist_m}(c), we found that the main CME with higher mass can more easily drive an enhanced type II radio burst than the lower mass main CME (up to about 60\% with high mass). Among all interacting CME pairs listed in this study, none of the main CME with log mass below 15.5 can generate any SEP event and the signature of radio enhancement. The en-type-II pre-CMEs also mostly have higher mass than the noen-type-II pre-CMEs (log mass: average 15.68 vs. 15.21, median 15.71 vs. 15.30) (Figure~\ref{Fig6.hist_m}(d,e)). The proportion of en-type-II events tends to increase almost linearly when the pre-CME mass increases, shown in Figure~\ref{Fig6.hist_m}(f).
From this figure, it's found that all SEP events (both large and small) are associated with the  interacting CME pairs with high mass (e.g. log mass of main CME $>\sim15.5$, pre-CME $>\sim14.5$), no matter if there is radio enhancement or not. The possible interpretation is that more massive CME might also be faster to drive a stronger shock to accelerate particles. It is also possible that interaction of two massive CMEs might generate a large SEP event accompanied by enhanced type II radio burst more easily through some mechanisms, e.g. twin-CME scenario.

From the analysis in Figure~\ref{Fig7.hist_e}(left four panels), we can see that the kinetic energy $E_k$ of both the main CMEs and the preceding CMEs is higher in group 1 than that in group 2, with average(median) log $E_k$ (unit:log(erg)) 31.93(31.87) to 30.93(30.97) for main CMEs and 30.48(30.54) to 29.55(29.59) for pre-CMEs. All(large) SEP events are accelerated by the main CMEs with log $E_k$ larger than 30.83(31.40), and by preceding CMEs with log $E_k$ larger than 28.887(28.892). However, the noen-type-II events can only lead to a few small SEP events with high energetic CMEs (e.g. log $(E_k)\geq30$ for the main CME and $\geq28$ for the pre-CME, shown in the middle two panels of
Figure~\ref{Fig7.hist_e}). With the kinetic energy of the main and preceding CMEs increasing, the proportion of events with type II radio enhancement also increase, as shown in the right two panels of Figure~\ref{Fig7.hist_e}. This result suggests that the type II radio burst enhancement and the production of large SEPs can be generated in favor of {CME interactions} with high kinetic energies.

   \begin{figure}[htb]
   \centering
   \includegraphics[width=0.8\textwidth, angle=0]{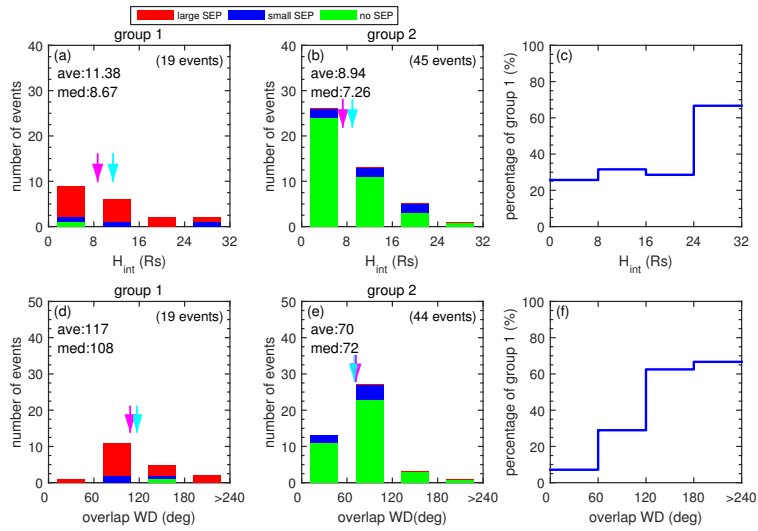}
   \caption{Same as Fig.~\ref{Fig4.hist_v}, but for the intersection height ($H_{int}$) and the overlap WD of CME pairs. }
   \label{Fig8.hist_h}
   \end{figure}

In Figure~\ref{Fig8.hist_h}(top panels), the average heliocentric distance where the leading-edge trajectories of the main CMEs intersecting that of the preceding CMEs is shown. This height is typically about 11.38$R_s$(median 8.67$R_s$) for en-type-II events and 8.94$R_s$(median 7.26$R_s$) for noen-type-II events. Since CMEs have finite thickness, the interaction of the CME pair must start before the intersection of the leading-edge trajectories.
So the release time of SEP near the Sun is earlier than this leading-edge intersection.
We do not find that the interaction height of the two CMEs can be used as a key factor to identify whether a pair has an enhancement of type II or not. According to the CPA and the WD values of the main and preceding CMEs from CDAW, we calculated and examined the overlap WD between two CMEs (see Figure~\ref{Fig8.hist_h}(bottom panels)). The overlap WD of en-type-II events is typically $\sim117^\circ$(median $108^\circ$) and  larger, comparing to the noen-type-II events $\sim70^\circ(72^\circ)$. The percentage of the en-type-II events in each bin of overlap WD tends to positively correlate with the overlap WD (Figure~\ref{Fig8.hist_h}(f)), which indicates that the interacting CME pairs having larger overlap WD (e.g. $\geq120^\circ$) will be more likely (e.g. $\sim60\%$ possibility) to lead to enhanced type II radio bursts. The large scatter of intersection heights and overlap WDs in SEP events in this figure also implies that the intersection height and the overlap WD of CME pair do not seem to be deciding factors for radio enhancement and SEP generation.

\subsection{SEP intensity dependence on the interaction of CMEs}
\label{sec.interact}
Since the {CME interaction} associated with enhanced type II radio bursts can generate SEP events more easily than that without type II or enhancement, what properties of CMEs' interaction affect on the intensities and occurrence of SEP events?

   \begin{figure}[htb]
   \centering
   \includegraphics[width=0.45\textwidth, angle=0]{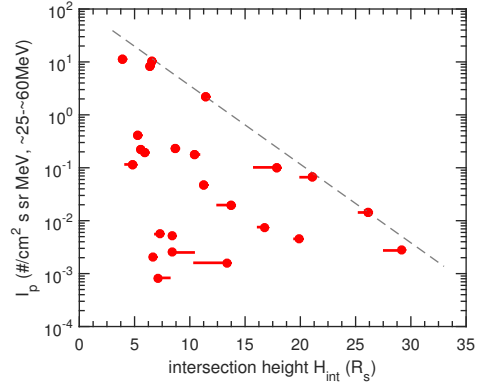}
   \caption{The intensity ($I_p$) of SEP events as a function of the intersection height $H_{int}$ of two CMEs leading-edge trajectories.
   %Red points denote the large SEP events, and blue ones denote the small SEP events.
   The dash line indicates the upper limit of the SEP intensities.}
   \label{Fig9.int_h}
   \end{figure}

Figure~\ref{Fig9.int_h} shows the correlation between the SEP intensity $I_p$ and the intersection height $H_{int}$ of main CME leading-edge trajectory overtaking the preceding CME in the FOV of SOHO/LASCO.
%The red points denote the large SEP events, and the blue points denote the small SEP events, separated by the horizontal dotted line.
As shown in the figure, the intensity indicates a distinctive upper limit, denoted by the dashed line, and correlate with the intersection height negatively. Also, most of SEP events (18/24,75\%, large events 12/15,80\%) are associated with the CME pairs having interaction height smaller than $15R_s$. A preceding CME captured by the main CME propagating at relative low height tend to drive a strong shock that more likely generates a large SEP event.

%, suggesting that energetic particles in a majority of SEP event are accelerated in a relative low height.
%From this figure, we can also see that the intensity of large SEP events correlates negatively with the intersection height with a correlation coefficient $R=-0.62$($SE=0.22$). For all SEP events, the correlation becomes weak ($R=-0.38,SE=0.20$).
%And it is suggested that a fast CME intersecting the preceding CME in the lower height can accelerate particles more likely to larger SEP event. However, we must
Note that the actual interaction of two CMEs must start below this intersection height of leading-edge trajectories because the CMEs have a finite thickness. The acceleration and detection of SEPs is controlled also by many other aspects, such as shock strength \citep[e.g.][]{Kahler96, Shen.etal07}, seed level \citep[e.g.][]{Kahler01, Ding.etal15}, magnetic field connection \citep[e.g.][]{Reames.etal96, Gopalswamy.etal05b}, so the scatter of SEP intensity as a function of intersection height is also very large.
%However, we must note that some full interactions occurred in the low height still can not produce the distinguishable SEP enhancement and radio enhancement.

   \begin{figure}[htb]
   \centering
   \includegraphics[width=0.8\textwidth, angle=0]{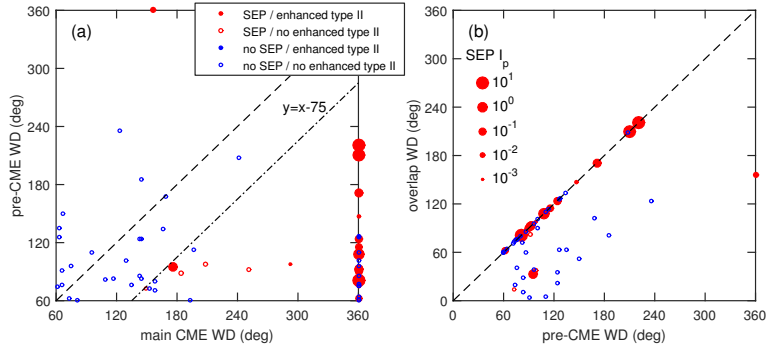}
   \caption{(a) Preceding CME WD vs. main CME WD; (b) Overlap WD vs. preceding CME WD.  Here, the filled circles indicate the events with enhanced type II, while the circles indicate the events without type II radio emissions or enhancement. Red color indicates the SEP-associated events, while blue color indicates the non-SEP events.
   The dash lines indicate equal values, and the dot-dash line shows a downward shift 75 from the dash line. }
   \label{Fig10.4d_wd}
   \end{figure}

In the process of CME interaction, the main CME may sweep partially or entirely over the body of the preceding CME. Perhaps a question one can ask is that does the relative width and the overlap WD of two interacting CMEs have some effect on the generation or intensity of SEP event. To answer this question, we tried to examine the correlations of WD and the overlap WD between main CME and pre-CME, shown in Figure~\ref{Fig10.4d_wd}. All SEP events but one (event 49) have been associated with the main CMEs wider (even about 75 degree larger, denoted by dot-dash line) than their corresponding preceding CMEs (see from panel (a)).
Perhaps the fact remains that most of large SEP events are accelerated by halo main CMEs.
The overlap WD of two CMEs vs. pre-CME WD is indicated in panel (b). It's interesting to find that the overlap WD of CME pairs in almost all SEP events, except four events, are close to the pre-CME WD,
which suggests that the main CME plow completely into the preceding CME (in the plane-of-sky). When this happens, SEP events seems to be more easily generated, especially large SEP events.
 Meanwhile, except two events, all en-type-II events are also associated with this type of CME pairs.
%However, these three exceptional events are labeled as the SEP events. Our result suggested that the interaction that the fast and wide CME completely overtaking the slow and narrow pre-CME usually can lead to the radio enhancement and SEP events more easily than the case with other manners such as narrow main CME, wide pre-CME, partial sweeping, and so on.

\section{Discussion and conclusion}

In this paper, we focused on 64 interacting CME pairs, and investigated what properties of the main and preceding CMEs best correlate with the enhancement of type II radio burst and whether the presence or absence of such enhancement is related to SEP events.
%including their interaction can effect distinctly on the acceleration of SEPs.
Various properties, such as CME speed, angular width, mass, kinetic energy, intersection height, and overlap WD, were examined in detail.

We approximated a comparative flux threshold of large SEP event to the value of 0.0114(0.01)/($cm^2ssr$MeV) for the observations of STEREO-A/B HET (SOHO/EPHIN) instrument in the energy range of $\sim25-\sim60$MeV, by which a large SEP event is defined equivalently to the event identified usually using $>10$pfu at $>10$MeV in GOES observations.
%The intensities of all large SEP events show positive correlations with the speed, kinetic energy and the mass of the main CMEs, but no association with the preceding CMEs. However, some properties of the main CMEs and the preceding CMEs may control the generation of (large) SEP events during the interaction of two CMEs.

For events in this sample, a vast majority of SEP events, including all individual large events,
 occur when the main CME overtakes the preceding CME and the accompanying type II radio burst shows enhancement. In contrast, only a few small SEP events occur for CME pairs without type II or enhancement. The en-type-II SEP events usually have wide longitudinal distribution, comparing to the noen-type-II SEP events. In all 64 interacting CME pairs, the probability of the SEPs occurrence for en-type-II events is higher than that of noen-type-II events.
It suggests that the presence of type-II radio enhancement can be used as a distinct signature of whether the interaction of CME pairs can produce a SEP event or not, especially large SEP event.
 We suggested that the presence of the signature of enhanced type II radio may be treated as a discriminator between SEP-rich and SEP-poor {CME interactions}. %\red{which is important for space weather.}

The statistical results show that the speed, WD, mass and kinetic energy of both main CMEs and preceding CMEs positively correlate with the probability of the presence of type II radio burst enhancement {during CME interactions}. The en-type-II events usually have higher speed, WD, mass and kinetic energy than the noen-type-II cases. These features imply that the main and preceding CMEs are more intense and energetic, which can more easily drive a stronger shock signified by type II radio bursts and enhancement.
%If enough seed particles can be provided, the shock is thereby expected to accelerate particles to high energy and generate large SEP event easily.

In our study, the intersection height and the overlap WD are roughly used to quantify the extent of the interaction of the two CMEs. The intersection height seems to show no distinct difference between the presence and absence of radio enhancement. But the intensity of SEP events was found to correlate inversely with the intersection height. This result indicates that if two CMEs interact in the lower corona during their propagation, due to perhaps a higher speed of the main CME, they can produce larger SEP events more easily.
%, which agrees with the fact that many large SEP or GLE events were accelerated and released at very low height \citep[e.g.][]{Reames09, Gopalswamy.etal12a, Ding.etal15}.
The overlap WD of en-type-II events is obviously larger than noen-type-II events. The portion of en-type-II events increases when the overlap WD becomes larger. All but one en-type-II events are associated with main CMEs having wider WD than that of the preceding CMEs.
However, it must be pointed out that most of main CMEs with radio enhancement are halos.
The result also shows that most of SEP events (20/24) are accelerated by the main CMEs widely overtaking the preceding CMEs or with largest overlap in the plane-of-sky. A possible interpretation may be that when {a} fast and wide CME widely sweeps up {a narrower and slower} preceding CME, particle acceleration of the shock can become more efficiently either because of the enhanced seed particles injected into the shock surface or because of the trapping and acceleration of the energetic particles in the closed flux loop of pre-CMEs intersecting with the shock surface.
We therefore suggested that if an energetic fast and wide CME overtakes its preceding CME fully in the low height, with the presence of enhanced type II radio emissions, then the CME pair generally can generate a high intensity SEP event. These results can help us to further understand the relationship between CME interaction and the large SEP events, and the mechanism of large SEP event that triggered by CME-driven shock.

\begin{acknowledgements}

We acknowledge the use of the online CME catalog CDAW (cdaw.gsfc.nasa.gov/CME\_list), SOHO, STEREO, GOES, Wind/WAVES and ground radio stations (Learmonth, Palehua, San-vito, Sagamore-hill) data provided through the indicated websites. This work is supported at NUIST by NSFC grants (Nos. U1731105, 41304150), Natural Science Foundation of Jiangsu Province of China (No. BK20171456), and sponsored also by Qing Lan Project of Jiangsu Province for L.G. Ding (2016). L. Feng is supported by NSFC grants 11522328, 11473070, 11427803, and also acknowledges the Youth Innovation Promotion Association and the specialized research fund from the State Key Laboratory of Space Weather for financial support.
\end{acknowledgements}

\bibliographystyle{raa}
%\bibliography{ref2017}



%   \begin{figure}[htb]
%   \centering
%   \includegraphics[width=0.8\textwidth, angle=0]{fig10.4d_m.eps}
%   \caption{(a) Height of two CMEs intersection vs. the mass of main CME; (b) Mass of preceding CME vs. that of main CME. Here, the filled circles present the type II enhanced events, while the circles present the no radio enhancement events. Red color indicates the SEP-associated events, while blue color indicates the SEP-poor events. }
%   \label{Fig10.4d_m}
%   \end{figure}

%

\end{document}